\documentclass[12pt]{iopart}
\usepackage{epsfig}

\begin{document}

\title[Atomic Fountains for Timing Applications]{Evaluation of Long Term Performance of Continuously Running Atomic Fountains}

\author{Steven Peil, James~L.~Hanssen, Thomas~B.~Swanson, Jennifer~Taylor, Christopher~R.~Ekstrom}

\address{United States Naval Observatory, Washington, D.C.  20392}
\ead{steven.peil@usno.navy.mil}

\begin{abstract}
An ensemble of rubidium atomic fountain clocks has been put into operation at the U.S. Naval Observatory (USNO). These fountains are used as continuous clocks in the manner of commercial cesium beams and hydrogen masers for the purpose of improved timing applications.  Four fountains have been in operation for more than two years and are included in the ensemble used to generate the USNO master clock.  Individual fountain performance is characterized by a white-frequency noise level below $2\times 10^{-13}$ and fractional-frequency stability routinely reaching the low $10^{-16}$s.  The highest performing pair of fountains exhibits stability consistent with each fountain integrating as white frequency noise, with Allan deviation surpassing $6\times 10^{-17}$ at $10^7$~s, and with no relative drift between the fountains at the level of $7.5 \times 10^{-19}$/day.  As an ensemble, the fountains generate a timescale with white-frequency noise level of $1\times 10^{-13}$ and long-term frequency stability consistent with zero drift relative to the world's primary standards at $1 \times 10^{-18}$/day.  The rubidium fountains are reported to the BIPM as continuously running clocks, as opposed to secondary standards, the only cold-atom clocks so reported.  Here we further characterize the performance of the individual fountains and the ensemble during the first two years in an operational environment, presenting the first look at long-term continuous behavior of fountain clocks.
\end{abstract}

\maketitle


\section{Introduction: Beyond Cesiums and Masers}

The performance of atomic clocks has improved more than seven orders of magnitude since the first cesium clock was demonstrated by Essen and Parry over 50 years ago~\cite{essen, aluminum}.  In the past two decades, improvement has been driven by several major scientific developments.  Most recently, optical frequency combs, narrow line width lasers, and methods of confining atoms and ions without perturbing the clock frequency have led to optical clocks, with reference frequencies and corresponding $Q$s orders of magnitude higher than previous systems.  Prior to optical clocks, the ground breaking techniques of laser cooling and trapping of atomic gases brought about dramatic improvements in atomic clock performance by enabling the creation of ultra-cold samples of atoms, providing longer interaction times and reduced systematic uncertainties.  The first clocks based on cold atoms were atomic fountains, which are in some sense the natural progression of atomic beam clocks. Atomic fountains are the basis of contemporary primary frequency standards, in which the microwave frequency of the hyperfine transition in cesium-133 is used as the reference for a clock~\cite{fountain}.

As the most precise and accurate frequency standards, cold-atom clocks (optical or microwave) find application in frequency metrology~\cite{aluminum}, precise tests of fundamental symmetries~\cite{LPI,LLI}, and as test beds for quantum scattering~\cite{gibble}, entanglement~\cite{entangle} and information processing~\cite{wineland}. However, the application of atomic frequency standards as clocks for timing applications is still dominated by clocks based on older but more mature technology, with far greater levels of engineering and evaluation.  The workhorses of the timing community, commercial cesium beams and hydrogen masers, which do not incorporate any laser technology, run continuously and require little maintenance and upkeep, often running for years without requiring service. International Atomic Time (TAI) utilizes data from hundreds of these clocks, and many timing labs rely exclusively on commercial cesiums and hydrogen masers for their local ensembles.

It is to be expected that the best optical clocks will require significant engineering before being compatible with user-free operation in a timing ensemble. (In fact, many of the best accuracy and stability evaluations carried out with optical clocks use measurements made at optical frequencies, without dividing the signal down to RF, which would be a requirement for most timing applications~\cite{10-18}.) On the other hand, atomic fountains are being used much more regularly for timing applications. As primary frequency standards, cesium fountains have been contributing to TAI for close to a decade, with the number and precision of reports increasing over time, and recently LNE-SYRTE began reporting a rubidium fountain to the BIPM as a secondary standard~\cite{SYRTE_sec}. And many timing labs are relying more on fountain clocks for their local timescales, with atomic fountains in use or under development in at least 10 nations worldwide.

Even with this increased use, the role of fountain clocks for timing has typically been different from the commercial clocks that dominate (in number) timing ensembles.  Serving as primary frequency standards, the fountains provide calibrations of the frequencies of other clocks.  In this role, they do not necessarily need to run continuously for months or years at a time without user intervention, though some have achieved close to continuous operation~\cite{PTB_timescale, SYRTE_trans}.  And for a local timescale, it can suffice to use only a single primary standard, which ensures a constant long-term frequency, at least on the time frame of the accuracy evaluations.  For example, the PTB has produced an excellent timescale based on a single cesium fountain~\cite{PTB_timescale}.

Still, there are advantages to having multiple clocks.  More clocks translate to more robustness and reliability.  And it is sometimes the case that a fountain clock is characterized by comparing to a second one.  For these reasons, most institutions with fountains have or are developing at least two; LNE-SYRTE has three and NPL is working on a third (in each case one of the three fountains uses rubidium).  Additional advantages of multiple clocks, particularly for timekeeping, are related to reducing noise. The white-frequency-noise level of an ensemble with $N$ clocks decreases as $\sqrt{N}$. And since continuously running atomic clocks exhibit non-stationary behavior, timing ensembles based on such clocks require multiple devices to optimize performance.

The U.S. Naval Observatory (USNO), the largest contributor of clock data to the BIPM, one of the official sources of time for the United States and the official time and frequency reference for critical infrastructure such as GPS, has designed and built an ensemble of atomic fountain clocks that has been in operation for more than 2.5 years.  The ensemble consists of four rubidium fountains housed in Washington, DC~\cite{amc}. These clocks run continuously, with little operator intervention required, and are used to enhance the local timing ensemble, contributing to the USNO master clock in a manner similar to commercial cesium beams and hydrogen masers. The specific role of the fountains is to provide a superior long-term frequency reference, in the past solely the job of the commercial cesiums, while the short-term reference continues to be provided by an ensemble of hydrogen masers.

All four rubidium fountains have been regularly reported to the BIPM for over two years as continuously running clocks, the first cold-atom clocks so reported. Here we present the performance of these clocks over the first 2 years of operation, the first characterization of continuously running atomic fountains over such time frames.

\section{Overview of Fountain Operation}

The four fountains in operation at USNO are designated NRF2, NRF3, NRF4 and NRF5.  The systems were built in two generations, according to slightly different designs~\cite{usno1,usno2,usno3}.  All four fountains began continuous operation in a dedicated clock facility in March 2011. Since then, some of the improvements in design used for the second generation, NRF4 and NRF5, have been incorporated as retrofits to the first generation, NRF2 and NRF3.  Additionally, both routine and unexpected maintenance were required on occasion.  Since December of 2011, fountain data for all four devices have been reported without interruption to the BIPM, as clock type 93.  In May of 2012, after the traditional evaluation period, the fountains were weighted in TAI for the first time, and all four clocks have received maximum weight in every evaluation since.

Each fountain has a local oscillator (LO) comprised of a quartz crystal phase-locked to a dedicated maser, which serves as the reference for the 6.8~GHz drive for rubidium spectroscopy as well as the reference for a precise frequency synthesizer, an auxiliary output generator (AOG).  Every 20 seconds, the average relative frequency of the fountain and maser is used to steer the AOG output, providing a continuous nominal 5~MHz signal that reflects the fountain frequency and is measured against the USNO master clock and other clocks in the ensemble.

During each fountain cycle, auxiliary data such as the number of atoms participating in the frequency measurement and the fraction of atoms making the transition between clock states are used to determine whether the fountain is operating normally. If these data fall outside of a predetermined healthy range, the fountain automatically switches to holdover mode, in which the maser serves as a flywheel; in this case, the control computer applies a steer that is the median frequency difference between the fountain and maser over the previous hour.  The same holdover steer is applied until the auxiliary data reflect healthy values and normal operation resumes.

The robustness of the fountains and the ability to rely on a maser as a flywheel have resulted in a high uptime over the 2 year period.  The percentage of time that each fountain has generated a good, steered output is 99.2\%, 99.7\%, 98.5\% and 100\% for NRF2 through NRF5.  A ``good output'' can include brief intervals where the system is in holdover; so long as the interval corresponds to an averaging time for which the relative frequency variations of the maser and fountain are still white ({\em i.e.\@} the frequency fluctuations are still limited by fountain performance). Periods in holdover long enough that the steered output reflects maser behavior is treated as ``down time'' and is not included in analysis of fountain behavior; it is handled in the data analysis by interpolating the frequency record and integrating to generate the phase record.

The quickest degradation of performance of the (nominal) 5~MHz output signal is the occurrence of a problem with the processor controlling fountain operation; if the steering algorithm is interrupted, the AOG will no longer be updated, and its output frequency will reflect a steer determined from 20~s of averaging (rather than the holdover value). Processor refresh is usually a simple process.  Problems that are more difficult to solve, but do not show up as quickly in the steered output, include shutter failure and diode laser aging.  Aside from these three problems that have arisen, routine maintenance is required on occasion.  This consists almost exclusively of occasional tweaks to optical alignment, sometimes not needed for months at a time. Very rarely an adjustment to the VCO on the reference maser or AOG is required to maintain phase lock.  There is no maintenance required for the laser frequency locks, which perform without incident.

\section{Clock Performance}

\subsection{White-Frequency Noise}

Perhaps in part due to the refinement in design, the highest performing fountains have been NRF4 and NRF5.  Each device exhibits a white-frequency noise level of $1.8 \times 10^{-13}$ and excellent long-term stability. (There is an obvious enhancement to the microwave chain which should improve the short-term stability to a level of about $1.4\times10^{-13}/\sqrt{\tau}$, which was demonstrated in an engineering prototype, NRF1.)

\begin{figure}
\centerline{\includegraphics[width=0.6\textwidth]{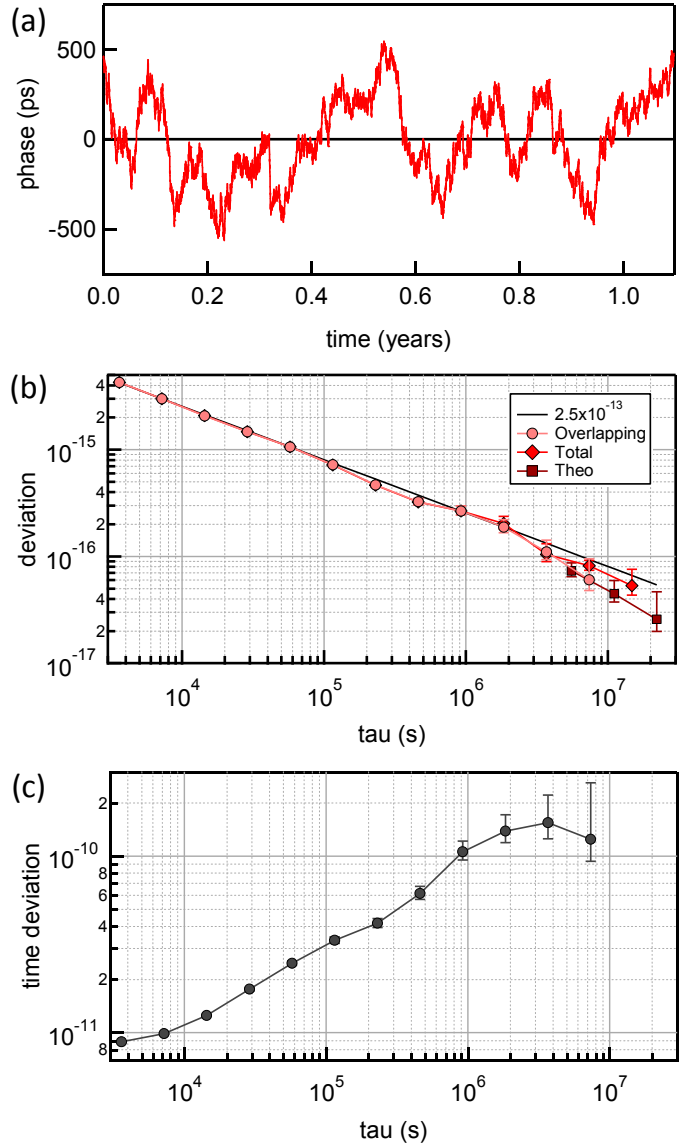}}
\caption{(Color online.) Performance of NRF4 vs NRF5 for 1.1~year interval. (a)~Residual difference in phase after removing a single frequency difference. (b)~Sigma-tau plot for data in (a).  Plot includes overlapping Allan deviation, total deviation, and Theo statistics.  The white-frequency noise reference line corresponds to an average level of $1.8\times 10^{-13}$ for each fountain.  (c)~Time deviation for data in (a).} \label{f.45good}
\end{figure}

In Fig.~\ref{f.45good}(a), the relative phase of NRF4 and NRF5 is plotted over an uninterrupted period of more than 1 year, after removing a relative frequency of $4.3\times 10^{-16}$ and a meaningless relative phase.   Residual peak-to-peak phase deviations stay within about 1~ns for the entire interval. A limit on the relative frequency drift can be obtained by fitting a line to the frequency, with the result that the relative drift is zero at the level of $7.5\times 10^{-19}$/day ($2.7\times10^{-16} $/year).

A sigma-tau plot, shown in (b), includes overlapping Allan deviation, total deviation and Theo statistics, along with a white-frequency noise reference line.  There is no indication of deviation from white-frequency noise behavior out to the total averaging time of more than 8 months. The average (in)stability for each fountain is consistent with an overlapping Allan deviation of $6.8\times 10^{-17}$ at $7\times 10^6$~s, a total deviation of $4.6\times 10^{-17}$ at $1.5 \times 10^7$~s and a Theo of $3.8\times 10^{-17}$ at $2.2 \times 10^7$~s (values correspond to the white-frequency reference level divided by $\sqrt{2}$, not to the data points which lie below this line). Figure~\ref{f.45good}~(c) shows the time deviation between the two fountains for the $\sim 1$~year interval; the average TDEV for each fountain is about 20~ps at a day.

It is likely that this represents record performance for microwave clocks.

\subsection{Non-stationary Behavior}

The data analyzed in Fig.~\ref{f.45good} correspond to an interval when both fountains exhibited excellent stability, which also coincided with a period of time during which there was little or no user intervention.  There were times during the two years when some final refinements were made to NRF4, upgrades were made to NRF2 and NRF3, and maintenance consisting of laser and shutter replacement were carried out on NRF2 and NRF3. When the entire two year data set is analyzed without any restrictions, and with no corrections or re-evaluations of the frequency across the service or upgrade, we see long-term behavior that deviates from white frequency noise.

\begin{figure}
\centerline{\includegraphics[width=0.6\textwidth]{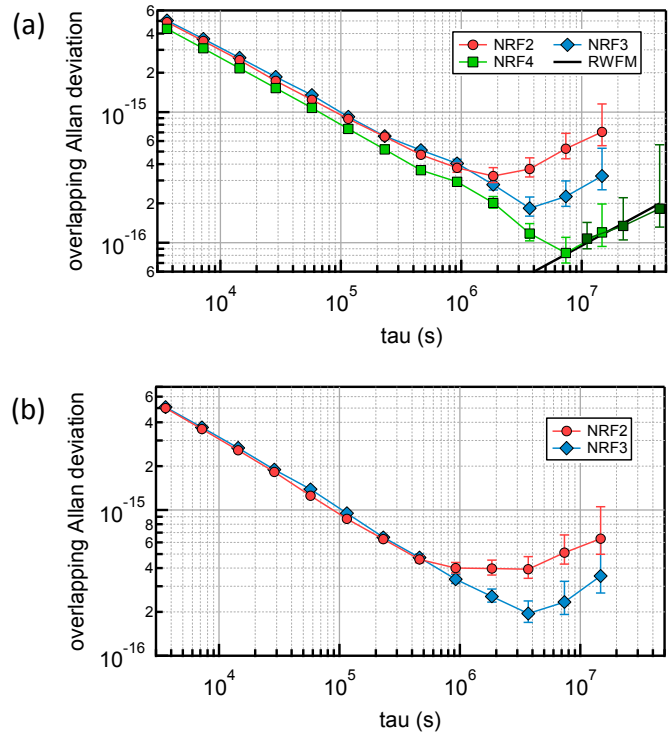}}
\caption{(Color online.) Summary of pair-wise fountain stabilities over first 2 years of operation. (a) Stabilities of NRF2, NRF3 and NRF4 measured against NRF5. RWFM--random walk frequency modulation. (b) Stabilities of NRF2 and NRF3 measured against NRF4.} \label{f.pairwise}
\end{figure}

Figure~\ref{f.pairwise} shows the overlapping Allan deviations for all pair-wise comparisons using the entire two-year data set.  The long-term behavior in all cases shows similar non-stationary behavior.  When the NRF4 vs NRF5 comparison in (a) is extended to longer averaging times using the Theo deviation (darker symbols), the most obvious fit is to random-walk frequency noise, with a $\sqrt{\tau}$ dependence.  The reference line in the figure corresponds to a random-walk level of $2\times 10^{-20} \sqrt{\tau}$ per fountain.  But, as we now show, the behavior leading to these stability plots is entirely consistent with discrete changes in frequency, at a rate that makes it impossible at this point to distinguish between inherent random-walk frequency noise and frequency changes arising from user intervention.

\begin{figure}
\centerline{\includegraphics[width=0.6\textwidth]{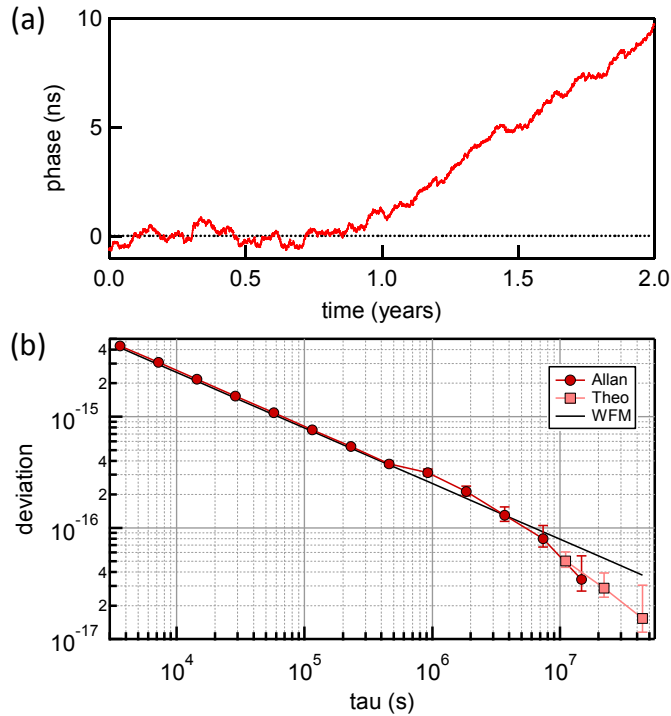}}
\caption{(Color online.) (a) Phase record for NRF4 vs NRF5, over the first two years of operation (from MJD 55637 to 56368).  (b) Overlapping Allan deviation and Theo for data set created when a single frequency is removed from the latter part of the phase record, whitening the data.  WFM--white frequency modulation.} \label{f.45white}
\end{figure}

In Fig.~\ref{f.45white}(a), we show the relative phase of NRF4 and NRF5 over the entire two year period.  The record of the relative phase shows two different linear regions with different slopes, indicating that the relative frequency changed (the data in Fig.~\ref{f.45good} correspond to the second frequency shown).  This bears out when we adjust the data starting from 0.9 years (MJD 55975) to the end by removing a single frequency, of size $< 3\times10^{-16}$, in order to make the last part of the frequency record match the first. The subsequent data look white, as illustrated by the sigma-tau plot in Fig.~\ref{f.45white}(b), indicating that the non-stationary behavior is indeed consistent with a single frequency change.

\begin{figure}
\centerline{\includegraphics[width=0.6\textwidth]{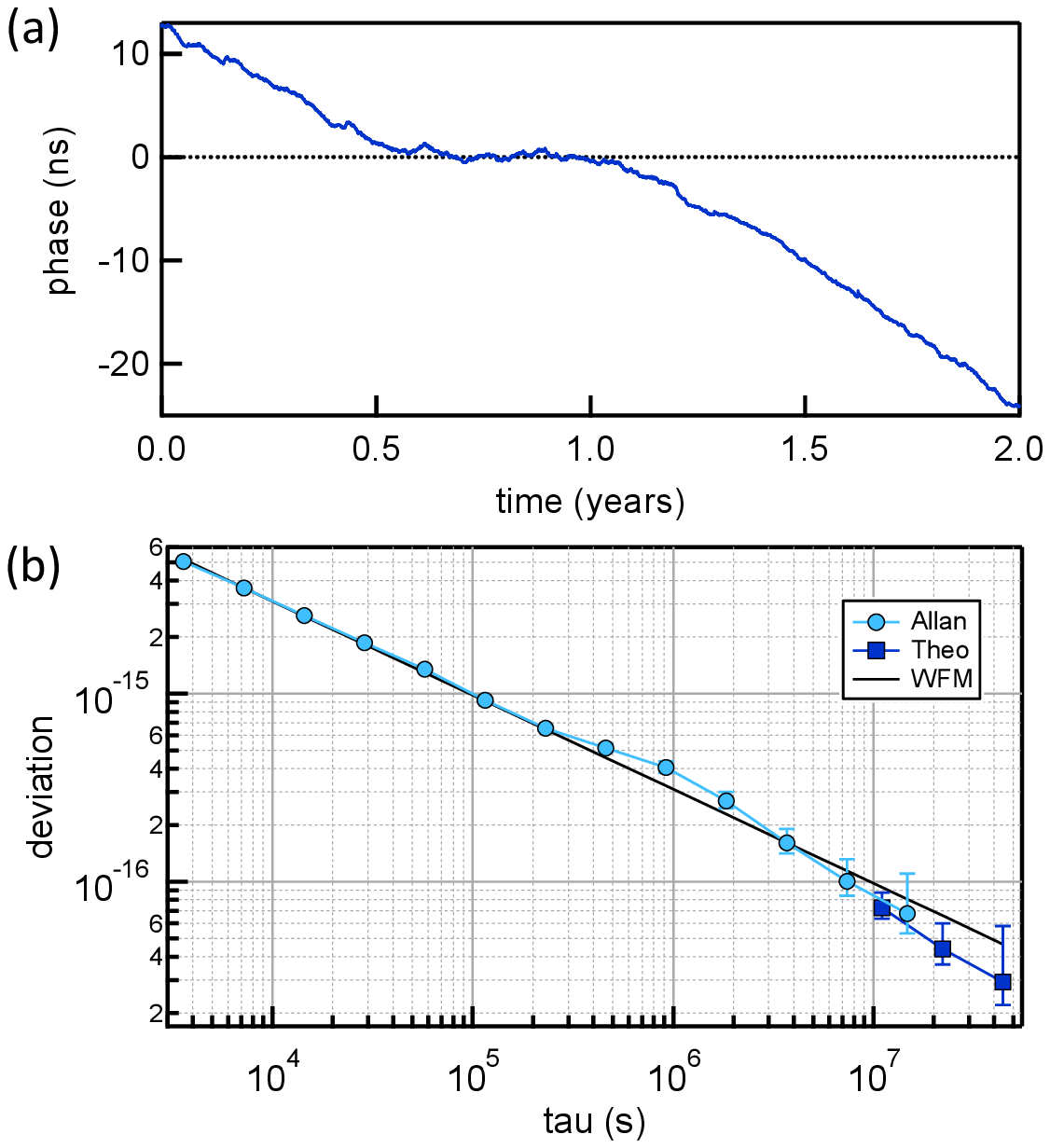}}
\caption{(Color online.) (a) Phase record for NRF3 vs NRF5.  (b) Overlapping Allan deviation and Theo for data set created when frequency adjustments are made to the data at two points.} \label{f.35white}
\end{figure}

Similar behavior was observed for NRF3. In Fig.~\ref{f.35white}(a), we show the relative phase of NRF3 and NRF5 over the entire two year period.  Here, the relative frequency appears to take on three different values. Adjusting the data at the two times when the frequency changed (0.6 yrs (MJD 55840) and 1.1 yrs (MJD 56030)), thereby adding two frequency adjustments, the resulting behavior is consistent with integration as white-frequency noise, as shown by the sigma-tau plot in Fig.~\ref{f.35white}(b). So for NRF3, NRF4 and NRF5, the long-term behavior under continuous operation is consistent with rare, discrete frequency changes, at an average rate of 1 every 2 years and an average size of $5.4\times10^{-16}$.

We can unambiguously assign the frequency changes in Fig.~\ref{f.35white} to NRF3, as they appear as well in comparisons with NRF4.  The frequency change in Fig.~\ref{f.45white} is more difficult to assign to one fountain or the other, because the other two fountains do not support a particular position. We suspect that the frequency change can be assigned to NRF4, however, because the time at which the change occurred coincides within the error bar to a time at which some finishing touches were completed on NRF4. Similarly, one of the changes in frequency exhibited in Fig.~\ref{f.35white}(a) occurred close in time to one of the upgrades made to bring the design of NRF3 closer to that of the second generation. It is impossible to conclude that these frequency changes were caused by user intervention, but it is a possibility, and it is something we intend to continue analyzing.

Almost all of the most likely sources of significant frequency shifts in a fountain have been ruled out as drivers of the observed behavior--the magnetic field is regularly measured and shows no correlated changes; light shifts do not seem to be a problem; the ambient temperature is monitored and is constant.  It is hard to conclusively rule out a microwave effect, particularly within the microwave chain.  It would be interesting to compare to primary standards, in terms of whether biases measured in accuracy evaluations change occasionally over time.

Because these frequency changes are rare, it is straightforward to recharacterize a fountain using the others as a reference.  This introduces the possibility of producing a paper fountain timescale that conceivably integrates as white-frequency noise to on order of $1\times 10^{-17}$.  We plan to consider this further.

The fourth fountain, NRF2, has exhibited the least stable behavior thus far.  It has shown discrete frequency changes, of larger size and at a higher rate than those discussed above, as well as intervals that more likely corresponded to drift.  Some of the problematic behavior in NRF2 does not correlate with episodes of significant intervention. We hope to be able to improve the performance of NRF2 and in the process identify the cause of at least some of this non-stationary behavior.

Finally, we point out that these changes in the output frequencies of the fountains are not observable using other clocks or timescales at USNO.  Neither the cesium ensemble, nor the maser ensemble, nor individual masers independently show that a fountain has changed frequency; these events are only observed with inter-fountain comparisons.

\section{Role in Local Timescale, UTC(USNO)}

For years the USNO master clock, the physical representation of UTC(USNO), has been generated by incorporating the long-term frequency stability of commercial cesium beams with the superior short-term stability of hydrogen masers.  Masers are well known to display time-variable drift and other behavior different from white-frequency noise after integrating to a level between low $10^{-15}$s and high $10^{-16}$s.  Each maser is re-characterized using the cesiums; the sooner that non-stationary behavior is detected, the better. One role of the rubidium fountains is to improve on the process of maser re-characterization.

\begin{figure}
\centerline{\includegraphics[width=0.6\textwidth]{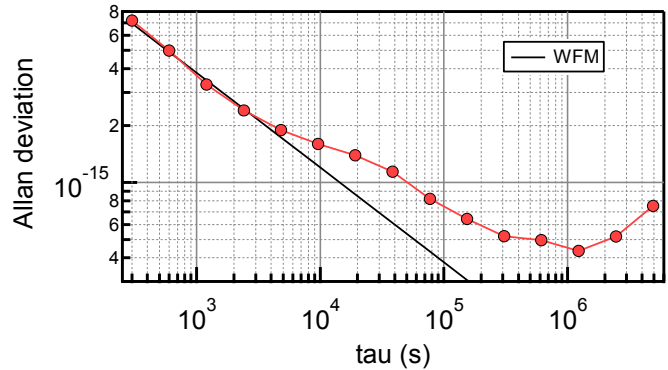}}
\caption{(Color online.) Overlapping Allan deviation of the average of the four fountains vs a hydrogen maser. The white-frequency noise reference line is at $1.2\times 10^{-13}$, corresponding to $1\times 10^{-13}$ for the fountain timescale and $6\times 10^{-14}$ for the maser.} \label{f.usno_timescale}
\end{figure}

In Fig.~\ref{f.usno_timescale} we show the Allan deviation of an exceptionally stable maser versus a crude fountain timescale, created by averaging the frequencies of the four rubidium fountains.  The short term relative stability of $1.2 \times 10^{-13}$ at 1~s is consistent with a white frequency noise level of $6\times 10^{-14}$ for the maser and $1\times 10^{-13}$ for the fountain timescale, as expected for four fountains with an average white-frequency noise of $2 \times 10^{-13}$. The white frequency noise level achieved with the USNO ensemble of about 70 commercial cesium beams is about $1\times 10^{-12}$.  So the fountain ensemble shows a factor of 10 improvement in short-term stability, which, in terms of averaging time, translates to a factor of 100 reduction.  Even a single fountain is more than 50 times better than the entire ensemble of cesiums.

As stated previously, it should be possible to improve the short-term stability of each fountain to about $1.4\times 10^{-13}$, making the fountain timescale white frequency noise level $7\times 10^{-14}$.  We expect that this would be the limit for our four-fountain ensemble. We have investigated using the same LO for a pair of fountains to enable reduction of the LO noise contribution.  Removing the entire LO noise contribution by differencing the fountain frequencies is possible for a paper clock and is useful for measuring atomic physics effects more quickly, but it cannot be incorporated into a physical output.  To generate a physical output with reduced LO noise, we implemented what we call ``cooperative steering'', where two fountains steer the same LO.  By synchronizing the load-measurement cycles so that the LO is always being measured and compensated for, the additional noise from not continually sampling the LO, the Dick effect, can be reduced~\cite{dick}. Because of the high quality of the LOs used for the rubidium fountains, the reduction in the Dick effect does not improve the stability greatly, while using a single LO for different fountains increases risk significantly. This increase in operational risk led us to abandon this technique.

As confidence in the reliability of the fountains has been gained, the clocks are now being used operationally as the long-term reference in the generation of UTC(USNO).  The specific algorithm for including the fountains as well as a quantitative assessment of their impact on UTC(USNO) is an area of ongoing research~\cite{koppang}.



\section{Stability vs Primaries Contributing to TAI}

\begin{figure}
\centerline{\includegraphics[width=0.6\textwidth]{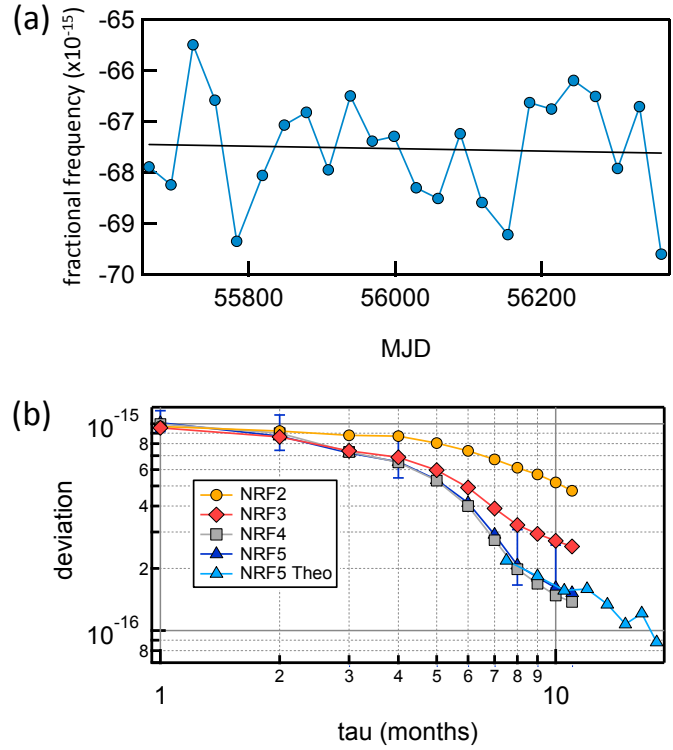}}
\caption{(Color online.) (a) Relative frequency of NRF5 and the average of the cesium fountains contributing to TAI for a particular interval. A line fit to the data gives a slope of $0.2 \pm 1.0 \times 10^{-18} $/per day, demonstrating that NRF5 exhibits zero drift with respect to the world's primaries at the level of $1.2 \times 10^{-18} $/per day. (b) Total deviation for each individual fountain vs the cesium fountains.  Theo deviation and sample error bars are included for NRF5.} \label{f.primaries}
\end{figure}

Eight cesium fountains worldwide contributed on some basis to TAI during the 2 years we are considering. Perhaps the best metric of long-term frequency stability of the USNO rubidium fountains is by comparing to these primary standards~\cite{primaries}.   We do this by using the USNO master clock, the physical realization of UTC(USNO), as a transfer oscillator to allow us to determine the frequency difference between a fountain and TAI, $\nu_{\rm Rb} - \nu_{\rm TAI}$.  Along with the values of the average frequency of the primaries compared to EAL, $\nu_{\rm Cs} - \nu_{\rm EAL}$, and the steers of EAL to TAI, $\nu_{\rm EAL} - \nu_{\rm TAI}$, which are published monthly by the BIPM in their Circular T report, the frequency of a fountain or fountain timescale compared to the primaries can be obtained.  Figure~\ref{f.primaries}(a) shows the monthly average of the frequency of NRF5 versus the average of the primaries. The line in the plot is a linear fit, $0.2 \pm 1.0 \times 10^{-18}$/day, showing that we can place a limit on the drift of NRF5 with respect to the primaries at the level of $1.2 \times 10^{-18} $/per day. In Fig.~\ref{f.primaries}(b), a sigma-tau plot shows the stability of each fountain versus the primaries.

\begin{figure}
\centerline{\includegraphics[width=0.6\textwidth]{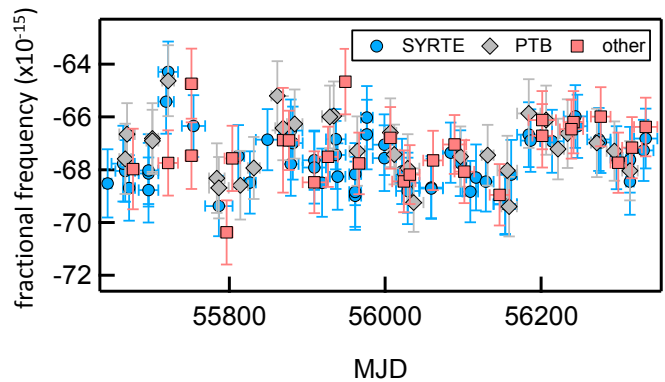}}
\caption{(Color online.) Relative frequency of NRF5 and each cesium fountain measurement reported to TAI.} \label{f.ind_primaries}
\end{figure}

Because of the continuous operation of our rubidium fountains, in addition to comparing to the monthly average of the primaries, we can obtain a measurement against each individual primary standard reported or against standards at a particular institution.  This allows us to analyze trends with respect to specific standards labs.  This is illustrated in Fig.~\ref{f.ind_primaries}, where the relative frequency of NRF5 and each primary standard reported to TAI over the 2 years is plotted. The (three) fountains at LNE-SYRTE and (the two) at PTB are differentiated from the other labs reporting (NIST, NPL and one report from NICT).

\section{Conclusion}

To summarize, the USNO has put into operation an ensemble of continuously running rubidium fountain clocks to incorporate into timing applications and timescale generation in a manner similar to hydrogen masers and commercial cesium beam clocks.  We have analyzed performance of the four fountains in the ensemble over the first two years of operation, demonstrating intervals reflecting record performance for microwave clocks and observing rare, discrete frequency changes. Comparison to the world's primary standards suggest that the rubidium fountains provide a stable, long-term frequency reference on par with any local reference.

\ack

We benefited from assistance from many members of the USNO Time Service Department, particularly Paul Koppang, Jim Skinner and Demetrios Matsakis.  Scott Crane played a significant role in the design and construction of the fountains. Atomic fountain development at USNO has been funded by ONR and SPAWAR.


\begin{thebibliography}{10}
\bibitem{primaries}
The primaries data set also includes CS1 and CS2 at PTB, which are beam standards.
\end{thebibliography}
\end{document}